\documentclass[aps,prl,preprint]{revtex4-2}

\usepackage{physics}
\usepackage{bm}
\usepackage{upgreek}
\usepackage{amssymb}
\usepackage{mathtools}
\usepackage{xcolor}
\definecolor{codegreen}{rgb}{0,0.6,0}
\usepackage{hyperref}
\hypersetup{
	colorlinks=true,
	linkcolor=blue,
	citecolor=codegreen
}
\newcommand{\RR}[1]{\mathrm{#1}}
\renewcommand{\eqref}[1]{equation (\ref{#1})}
\newcommand{\pref}[1]{(\ref{#1})}

\newcommand{\mean}[1]{\left< #1 \right>}

\newcommand{\vphi}{\bm{\upphi}}

\usepackage{scalerel}

\begin{document}
%opening
\title{A New Method for the Calculation of Functional and Path Integrals}
%Original title: New Method for the Calculation of Functional Integrals}
\author{Amos A. Hari}
\affiliation{Faculty of Mechanical Engineering, Technion - Israel Institute of Technology}
\author{Sefi Givli}
\affiliation{Faculty of Mechanical Engineering, Technion - Israel Institute of Technology}
\date{}

\begin{abstract} % abstract
	Functional integrals are central to modern theories ranging from quantum mechanics and statistical thermodynamics to biology, chemistry, and finance. In this work we present a new method for calculating functional integrals based on a finite-element formulation. This approach is far more robust, versatile, and powerful than existing methods, thus allowing for more sophisticated computations and the study of problems that could not previously be tackled. Importantly, existing procedures, element libraries and shape functions, which have been developed throughout the years in the context of engineering analysis and partial differential equations, may be directly employed for this purpose.
\end{abstract}

\maketitle

Functional (or path) integrals, are ubiquitous in a wide range of physical and mathematical problems, ranging from quantum mechanics to statistical thermodynamics through biology, chemistry, engineering and finance \cite{Feynman1942, feynman1948, FeynmannHibbs1965, pbcChap10, chemistryPRL, chemistryJCP, Reliability1, Reliability2, finance}.
Functional integrals generalize the notion of integration over \emph{vector spaces} to integration over \emph{function spaces}. Much like vector integration, $\int f(\vb{v})\dd{\vb{v}} $, where the integral is the sum of all volume elements in the integration space each weighted by an integrand \emph{function}, functional integration, $\int g[u]\mathcal{D}u $, is the summation process over all admissible functions each weighted by an integrand \emph{functional}.
Currently, the prevailing method for evaluating path-integrals is a "slicing method" which involves a naive discretization of the spatial (or temporal) space followed by summation over the function values at the discrete points \cite{Feynman1942, feynman1948, FeynmannHibbs1965}. In that approach, for example, derivatives of the state function with respect to the independent coordinates are approximated using finite-differences. The "slicing method" is straight forward and useful, however it is limited to 1-D systems in time or space, or to very simple and regular spatial geometries, such as rectangles. Further, even in 1-D, the application of certain boundary conditions, as well as constraints inside the domain, can be cumbersome. Thus, there is a clear need in a new, more sophisticated, method for the calculation of functional integrals.

In this paper, we propose a new approach for the calculation of functional integrals that is based on the finite-element (FE) formulation rather than the aforementioned na\"ive discretization. The FE method is a numerical approximation method predominantly used for solving partial differential equations (PDEs) in all fields of engineering, ranging from stress analysis and heat transfer in structures to electromagnetic scattering of objects and the design of photonic crystals \cite{hughes, bathe, magneticFE, scattering, photonicCrystals}. Driven by the growing need to tackle more complicated engineering problems, the FE method has become the standard numerical tool for engineering design and mechanical analysis, thanks to its generality, robustness and versatility.
Still, the use of the FE approach for evaluating functional integrals has been largely overlooked, practically limiting the computation of path integrals to 1-D or to very simple geometries. 

As we show below, applying the FE approach to functional integrals requires some technical care; however, the powerful formulation opens the door for more sophisticated computations and for the study of problems that could not previously be tackled. In that sense, it may be reminiscent of the revolution brought by the FE approach to partial differential equations, which has enabled solving complex engineering problems with complicated geometries and all types of boundary conditions or constraints. Further, although the method we propose is new, the underlying foundations are mature and well-established: Countless papers and textbooks have been published on the theory of FE \cite{hughes, bathe, zienkiewiczFE, whiteFE, raoFE, fractureFE}, and highly sophisticated computational schemes and software have been developed; all may readily be repurposed for our needs. For example, one may directly use the large established libraries of finite elements and associated shape-functions \cite{defelement}, or apply well-established meshing procedures and related software \cite{gmsh}. Perhaps the most important advantage that the FE formulation introduces to the computation of functional integrals is the preservation of the state functions as functional entities; this is unlike the existing approaches where the state functions are reduced to a finite set of discrete points. This property of the method allows for a more rigorous treatment and valuable insights. Moreover, even in simple cases, where the formulation does not provide fundamentally new results, we may still find ourselves appreciating the refreshing interpretations the method gives, or by the words of Richard P. Feynman \emph{there is a pleasure in recognizing old things from a new point of view} \cite{feynman1948}.

\paragraph{The Method.---}
%we need to use "LaTeX commands" for sections and not create our own typesetting style. The journal may later decide how it style the section titles. I assume you changed the title style so it take less space, but it was quite hard for me to navigate the text without a bold title. If this is even up to us then I suggest we use on the \paragraph{} command -- inline bold title.
%\emph{The method$.-$}
Consider a stochastic system whose (micro)state is described by the real valued function $ u(x)\in V $, where $ V $ is the space of all possible states, or admissible functions, of the system, and $ x $ is a point in a parameter space $ \Omega $. At the moment don't make any assumption on the class of continuity of $ V $.
Let the functional $ p[u] $, with $ p:V\rightarrow \mathbb{R} $, be the probability density corresponding to state $ u(x) $; and let $g[u]$ be some functional, $g:V\rightarrow\mathbb{R}$, dependent on the system's state.
The average of the state-dependent functional $g[u]$ is defined as the sum of $g[u]$ over all possible states of the system weighed by the respective probability $p[u]\mathcal{D}u$, hence expressed as a \emph{functional integral}
\begin{equation}\label{eq: func-int}
	\mean{g[u]} = \int_V g[u]p[u]\mathcal{D}u.
\end{equation}
The mathematical rigor for such operation is quite subtle, but the concept is well established, e.g., the Feynman path integral \cite{Feynman1942, feynman1948, FeynmannHibbs1965}; in that context, \eqref{eq: func-int} should be regarded as a formal way to express the functional averaging process.

We hereby present a new method for the calculation of functional integrals, based on a FE formulation.
As a first step, we approximate the function space $ V $ by introducing a subspace $ V^h\subset V $. The superscript $ h $ is called the \emph{mesh parameter} and it implies that the functions $ u^h\in V^h$ are associated with a \emph{mesh}, or a discretization, of the domain $ \Omega $. The mesh parameter, $ h $, is a measure for the size of the largest element in the mesh, therefore when $ h\rightarrow0 $ then $ V^h\rightarrow V $.
Following standard FE practice, we define $ u^h\in V^h $ as follows
\begin{equation}\label{eq: FE-approx}
	u(x) \approx u^h(x) = \sum_{A\in\eta} \phi_A(x) d_A + \sum_{A\in\eta_u}\phi_A(x)\bar{u}_A.
\end{equation}
Here $ \eta $ is the set of \emph{open nodes}, i.e. nodes where $ u $ is variable, and $ \eta_u $ is the set of \emph{closed nodes}, i.e. nodes where $ u $ is prescribed.
The functions $ \phi_A(x) $ are called FE shape functions; these are continuous functions that are related to the mesh, in the sense that each function $ \phi_A(x) $ gets the value of one at node $ A $ and vanishes in all elements that don't contain that node.
It is therefore apparent that the coefficients $ d_A $ and $ \bar{u}_A $ are the values of $ u $ at the respective node.
It is shown in FE theory that as $ h\rightarrow0 $ then $ u^h\rightarrow u $ and the approximation error is given by $ \norm{u-u^h} = Ch^\alpha $, where $ \alpha $ is the rate of convergence which depends on the type of FE shape functions used \cite{hughes}.

Once we are able to represent $ u $ by a finite number of degrees of freedom (DOFs), we may substitute the approximation $ u^h $ into the probability density and obtain $ p^h(\vb{d}) $. Note that $ p^h $ is a \emph{function} of the variables $ \vb{d}=\qty{d_A} $, rather than a \emph{functional}, and it is normalized such that $ \int p^h(\vb{d})\dd{\vb{d}}=1 $.
% I added a note about this subtelty. I kept finding myself wondering about it and then arriving again to the same conclusion. I think this needs to be said for this article to be a complete reference.
Similarly, we express $ g^h(\vb{d}) = g[u^h] $ as a function of $ \vb{d} $.
%Because instead of since.
Since $ \vb{d} $ uniquely determines $ u^h $ and $ g^h $, then the problem of finding $ \mean{g[u^h]} $ reduces to computing $ \mean{g^h(\vb{d})} $, where
\begin{equation}\label{eq: approx-int}
	\mean{g^h(\vb{d})} = \int_{\mathbb{R}^{N}} p^h(\vb{d}) g^h(\vb{d})\dd{\vb{d}}.
\end{equation}
Above, $ N $ is the number of DOFs (the cardinal number of $ \eta $) and $ \mathbb{R}^{N} $ is the space of all real vectors of length $ N $.
Equation \pref{eq: approx-int} is thus the FE approximation of the functional integral \pref{eq: func-int}, and the finite-dimensional integration can be carried out analytically (where possible) or numerically. The Markov-Chain Monte-Carlo method is especially appropriate for this task, because its rate of convergence is independent on $ N $ and it is particularly suitable for finding the statistical moments of complicated distributions \cite{Dunn}.

\begin{figure}
	\centering
	\includegraphics[width=\linewidth]{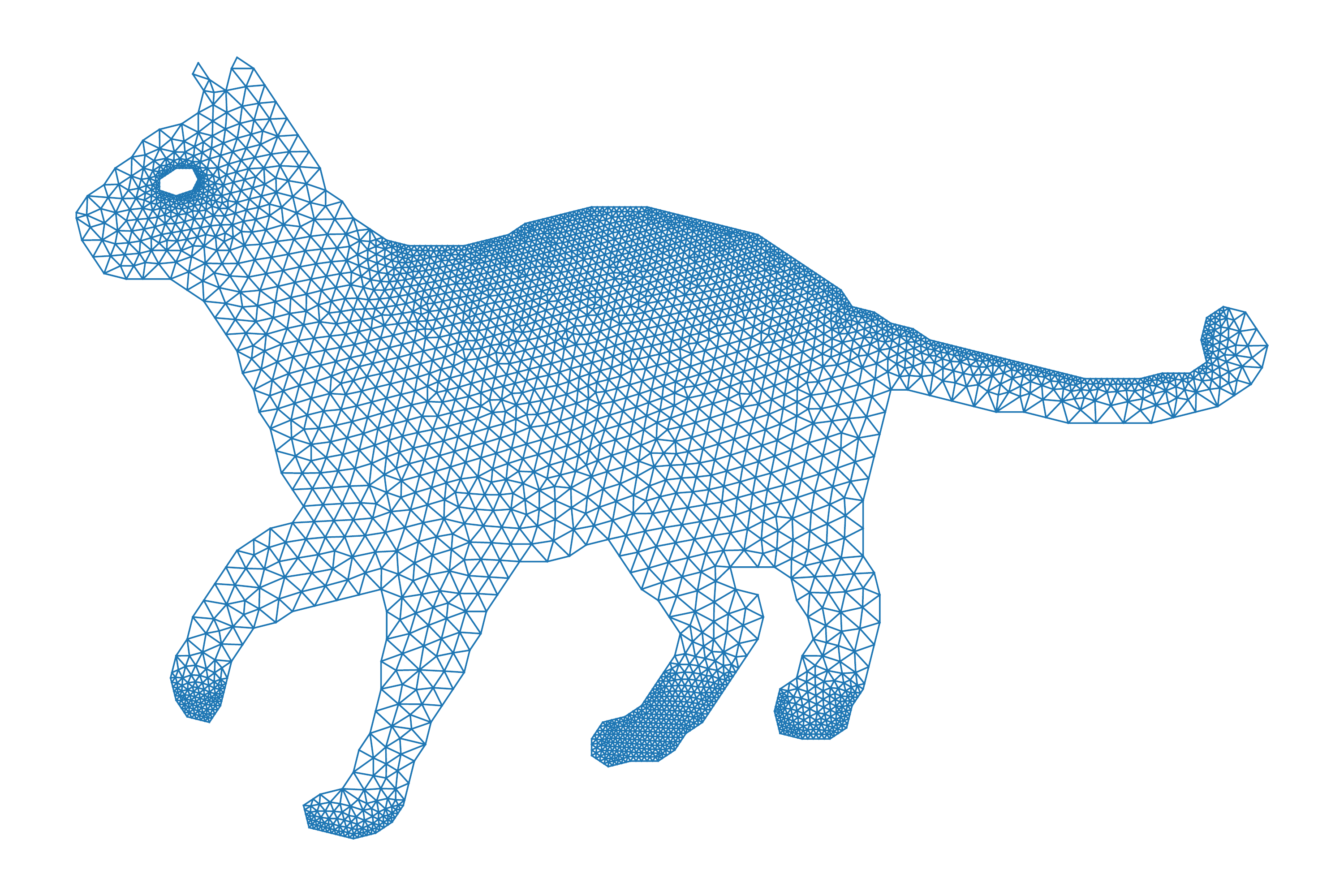}
	\caption{Example of a non-uniform mesh over a 2D domain in the shape of a cat's silhouette. The mesh was generated using the open-source Gmsh \cite{gmsh}.}
	\label{fig: treble_clef_mesh}
\end{figure}

Note that the method described here can be readily generalized for vector functions $ \vb{u}=\vb{u}(x)$ by writing equation \pref{eq: FE-approx} separately for each component of $ \vb{u} $. Further, the domain $ \Omega $ may be a simple 1-D interval, as in the case of the path integral in quantum mechanics where the coordinate $x$ above represents time, or be a multi-dimensional domain of any geometry, such as in the case of statistical thermodynamics of a 3-D body. 
Unfortunately, the "slicing method", commonly adopted for computing path integrals, cannot be applied to the latter. In terms of the proposed method, however, the only difference between these two cases is the use of different finite elements and corresponding shape functions; while in the 1-D case the elements are lines (or 1-D segments) and the shape functions are described in terms of one coordinate, in 2-D the elements have a 2-D geometry, such as triangles or quadrilaterals, and the corresponding shape functions are described using two coordinates. Similarly, if the domain $\Omega$ is three-dimensional, 3-D elements are used, etc.
It is noted that the use of a non-uniform mesh, where the domain $ \Omega $ is divided into elements of different sizes, is a standard practice of the FE method as illustrated in figure \ref{fig: treble_clef_mesh}. This allows, for example, to use of a finer mesh in regions where high accuracy is needed.
% edited the sentence to sound better
This feature is another important attribute of the versatile and powerful finite-element formulation. 
Finally, the physics of the problem dictates the number of DOFs at each node of the element. This is exemplified in the two examples below, where the first example involves one DOF at each node, while for the second example two DOFs are used at each node.

%\begin{figure}
%	\centering
%	\includegraphics[width=0.2\linewidth]{mesh example/treble_clef_mesh.pdf}
%	\caption{Examples}
%	\label{fig: treble_clef_mesh}
%\end{figure}

%\section{Examples}
%\subsection{String}
\paragraph{Example I: string.---}
%\emph{Example I: string$.-$}
Consider a string of length $ L $ and uniform tension $ \sigma $ with both ends held fixed at a horizontal level.
A lateral force $ f(x) $ is distributed along the string, and the entire system is submerged in a heat reservoir of temperature $ T $.
Let $ u(x)\in V $ describe the transverse displacement of the string at $ x\in\Omega=[0,L] $ and regard $ u $ as the state of the system. We would like to find, for example, the average state of the system.
The space $ V=\qty{u|u\in H^{1}(\Omega), u(0)=u(L)=0} $ 
is the set of all square-integrable functions over $\Omega$ with square-integrable first derivatives (Sobolev space) that admit the boundary conditions $u(0)=u(L)=\bar{u}=0$. 
The probability density corresponding to micro-state $ u $ is \cite{reif}
\begin{equation}\label{eq: Boltzmann-dist}
	p[u] = \frac{1}{Z}e^{-\beta E[u]},
\end{equation}
where the partition function, $ Z $, is a normalization constant, $ \beta=\qty(k_BT)^{-1} $, and $ E[u] $ is the energy functional
\begin{equation}\label{eq: string-energy}
	E[u] = \underbracket{\int_\Omega \frac{1}{2}\sigma u_{,x}^2\dd{\Omega}}_{\text{tension}} - \underbracket{\int_\Omega fu\dd{\Omega}}_{\text{load}},
\end{equation}
with $ u_{,x}=\pdv*{u}{x} $. 
Accordingly, the average state of the system is given by the functional integral
\begin{equation}\label{eq: string-mean}
	\mean{u} = \frac{1}{Z}\int_V u e^{-\beta E[u]} \mathcal{D}u.
\end{equation}
Following the method described above, we write the finite-element approximation as
\begin{equation}
	u\approx u^h = \sum_{A\in\eta} \phi_Ad_A + \sum_{A\in\eta_u} \phi_A\bar{u}_A,
\end{equation}
where, for simplicity, $ \phi_A(x) $ are linear FE shape functions \cite{hughes}, and thus $ u $ is approximated by $ u^h $ in a continuous piece-wise-linear manner.
Next, we substitute the FE approximation $ u^h $ into the energy functional $ E[u] $ so it becomes a function $ E^h(\vb{d})=E[u^h] $ of the variable $ \vb{d} $. This energy integral is calculated at the \emph{element level}.
Thus, define the vectors $ \vphi(x)=\qty{\phi_1^e(x), \phi_2^e(x)}^T $ of the element shape functions and $ \vb{d}^e=\qty{d_1^e, d_2^e}^T $ of the element state values, such that the state in each element is given by $ u^h = \vphi^T\vb{d}^e $.
Then, rewrite the integral \pref{eq: string-energy} as a sum of integrals over $ \Omega^e\subset\Omega $, the domain of the $ e $-th element.
The energy of that element is then
\begin{equation}\label{eq: string-element-energy}
	(E^h)^e = \frac{1}{2}\vb{d}^{eT}\vb{k}^e\vb{d}^e - \hat{\vb{f}}^{eT}\vb{m}^e\vb{d}^e,
\end{equation}
where $ \hat{\vb{f}}^e=\qty{\hat{f}^e_1,\hat{f}^e_2}^T $ is a vector whose components are the values of $ f $ at the element nodes, such that $ f\approx f^h=\vphi^T\hat{\vb{f}}^e $ describes the force inside the element.
The symmetric matrices $ \vb{k}^e=\sigma\int_{\Omega^e}\vphi_{,x}\vphi_{,x}^T\dd{\Omega} $ and $ \vb{m}^e=\int_{\Omega^e}\vphi\vphi^T\dd{\Omega} $ are respectively called the element stiffness and mass matrices and they are calculated in each element separately; however, for many elements, including the 1D element considered here, a formula for these matrices is readily found in the literature \cite{hughes, defelement}.
Next, define the matrix $ \vb{K} $, vector $ \vb{v} $ and scalar $ S $ such that the expression for the global energy becomes
\begin{equation}\label{eq: quadratic-energy}
	E^h(\vb{d})=\frac{1}{2}\vb{d}^T\vb{K}\vb{d}+(\vb{v}-\vb{F})^T\vb{d} + \frac{1}{2}S.
\end{equation}
Here, $ \vb{K} $ is the \emph{global} stiffness matrix, the vectors $ \vb{v} $ and $ \vb{F} $ are related to the contribution of prescribed displacements and of external loads, respectively, and the scalar $ S $ corresponds solely to the contribution of boundary condition to the energy.
This quadratic energy may now be substituted into \eqref{eq: Boltzmann-dist} so it becomes an off-centered Gaussian and thus $ \mean{\vb{d}} $ may be calculated analytically from \eqref{eq: approx-int} by considering the particular case of $g^h(\vb{d})=\vb{d} $ .

%\subsection{Beam}
\paragraph{Example II: Beam.---}
%\emph{Example II: beam$.-$}
Consider a stochastic system that is modeled by an Euler-Bernoulli beam of length $ L $ and uniform bending stiffness $ K_B $ with one end fixed and the other free.
A lateral load $ f(x) $ is distributed along the beam, and the entire system is submerged in a heat reservoir of temperature $ T $.
Let $ u(x)\in V $ describe the transverse deflection of the beam at $ x\in\Omega=[0,L] $ and regard $ u $ as the state of the system.
The space $ V=\qty{u|u\in H^{2}(\Omega), u(0)=0, u_x(0)=0} $ is the set of all square-integrable functions over $ \Omega $ with square-integrable first and second derivatives that admit the fixed boundary conditions $ u(0)=0,\ u_x(0)=0 $. Similar to the previous example, we want to find the average state of the system. The statistical distribution is given by \eqref{eq: Boltzmann-dist}, with the energy functional
\begin{equation}\label{eq: beam-energy}
	E[u] = \underbracket{\int_{\Omega} \frac{1}{2}K_B u_{,xx}^2\dd{\Omega}}_{\text{bending}} - \underbracket{\int_{\Omega} fu \dd{\Omega}}_{\text{load}}.
\end{equation}
The fundamental difference compared to the previous example is that for the energy \pref{eq: beam-energy} to be well defined, we demand stronger requirements on the class of continuity of $ u $, namely that $ V\subset H^2 $.
Accordingly, $ u^h\in V^h $ must satisfy these continuity requirements. To this end, we approximate $ u $ using the Hermite cubic shape functions $ \phi_{(Ai)} $ where $i=1,2$ \cite{hughes}. These shape functions and their derivative vanish everywhere, except in the elements that share node $ A $. Moreover, at node $ A $, the shape functions satisfy $ \phi_{(A1)}=\phi_{(A2),x}=1 $ and $ \phi_{(A2)}=\phi_{(A1),x}=0 $.
%they take the value of one if $ i=1 $ and their derivative takes the value of one if $ i=2 $.
This property allows dictating separately the displacements ($ u $) and rotations $ (u_{,x}) $ at the element nodes, thus
\begin{equation}
	u\approx u^h = \sum_{i=1}^{N_{\RR{ndof}}}\qty(\sum_{A\in\eta^{(i)}} \phi_{(Ai)}d_{(Ai)} + \sum_{A\in\eta_u^{(i)}}\phi_{(Ai)}\bar{u}_{(Ai)}).
\end{equation}
Here, $ N_{\RR{ndof}} $ is the number of nodal DOFs (in our case, $ N_{\RR{ndof}}=2 $). The values of $ u $ at the nodes are $ d_{(A1)} $ and the values of $ u_{,x} $ at the nodes are $ d_{(A2)} $, the sets $ \eta^{(i)} $ and $ \eta_u^{(i)} $ are defined as in \eqref{eq: FE-approx} but they include only nodes with open or closed $ i $-th DOF.
Define the vector $ \vphi(x)=\qty{\phi_{(11)}^e(x), \phi_{(12)}^e(x),\phi_{(21)}^e(x), \phi_{(22)}^e(x)}^T $ of the element shape functions and the vector $ \vb{d}^e=\qty{d_{(11)}^e, d_{(12)}^e,d_{(21)}^e, d_{(22)}^e}^T $ of the element state values, such that the state in each element is given by the cubic function $ u^h = \vphi^T\vb{d}^e $. The energy expression for the element is the same as in \eqref{eq: string-element-energy}, other than $ \vb{k}^e = K_B\int_{\Omega^e}\vphi_{,xx}\vphi_{,xx}^T\dd{\Omega} $; therefore the global energy also has the quadratic form of \eqref{eq: quadratic-energy} and $ \mean{\vb{d}} $ may be calculated analytically from \eqref{eq: approx-int}. 

%\subsection{Numerical Example}
\paragraph{Adhesion of elastic body to a rigid substrate: a numerical example.---}
%\emph{Adhesion of elastic body to a rigid substrate: a numerical example$.-$}
In what follows, we present numerical results obtained using the proposed finite-element formulation. The model considered is prototypical to phenomena such as detachment of biological cells, peeling of a thin film from a substrate, etc., and demonstrates how the formulation can be conveniently applied to complex systems composed of coupled linear and non-linear elements. We emphasize that while the model may be suitable for describing real phenomena, such as those mentioned above, it is presented here merely for demonstrating the proposed method; thus justification of the model and its assumptions are not further discussed.

\begin{figure}
	\centering
	\includegraphics[width=\linewidth]{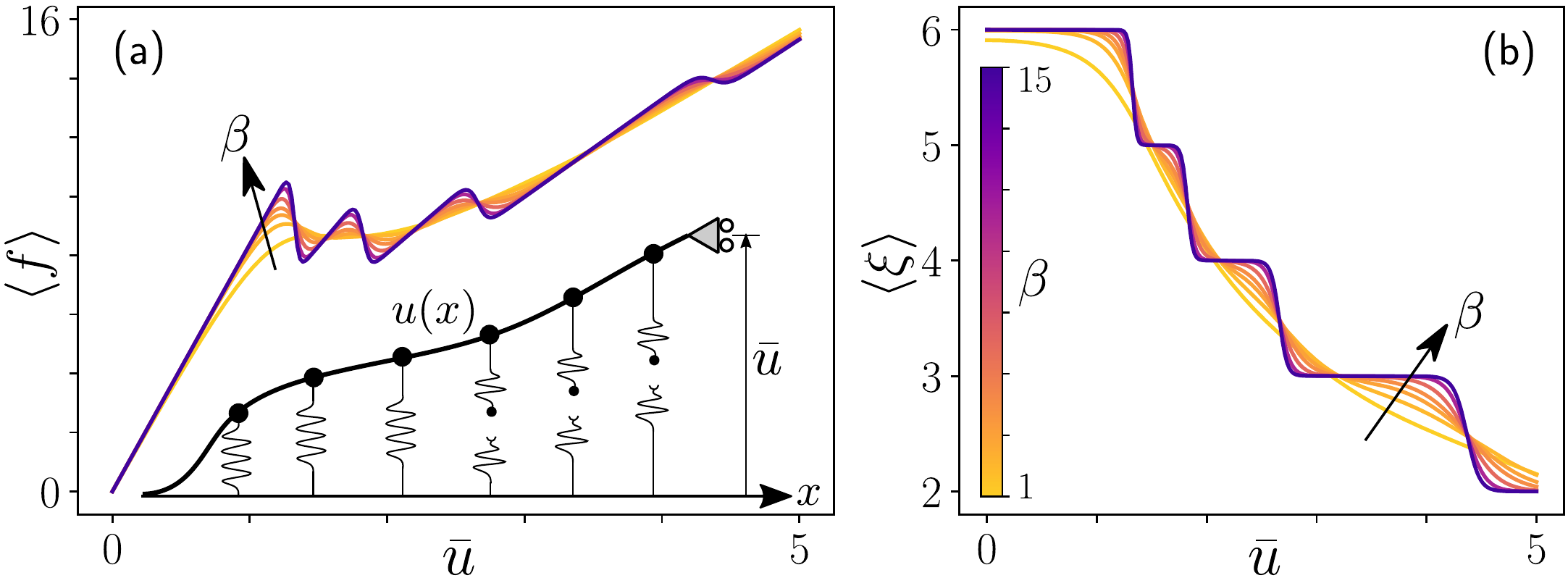}
	\caption{Relation between the end-displacement $ \bar{u} $ and (a) the mean force $ \mean{f} $, or (b) the mean number of attached bonds $ \mean{\xi} $, for various values of $\beta$. Inset: schematic illustration of the model.}
	\label{fig: force-disp}
\end{figure}

%\begin{figure}
%	\centering
%	\includegraphics[width=6cm]{NumExample_ver.png}
%	\caption{Relation between the end-displacement $ \bar{u} $ and (a) the mean force $ \mean{f} $, or (b) the mean number of attached bonds $ \mean{\xi} $, for various values of $\beta$. Inset: schematic illustration of the model. \red{clearly the vertical arrangement is clearer, but takes double the space. Perhaps we can change the font size for the horizontal arreangement such that it looks good} }
%	%\label{fig: force-disp}
%\end{figure}

%\begin{figure}
%	\centering
%	\includegraphics[width=\linewidth]{force_disp.pdf}
%	\caption{Relation between the end-displacement $ \bar{u} $ and correspoding mean force $ \mean{f} $ for various temperatures. Inset: schematic illustration of the model. \red{we need to combine the two figures into one figure (with "(a)" and "(b)") to save space. Also each of the figures will be smaller. Remove the title at the top of each fig, unite the x-axis for both. Font in the inset is not clear enough}}
%	\label{fig: force-disp}
%\end{figure}

%\begin{figure}
%	\centering
%	\includegraphics[width=\linewidth]{xi_disp.pdf}
%	\caption{(a) The mean number of attached bonds $ \mean{\xi} $ as function of the prescribed displacement $ \bar{u} $. As in figure \ref{fig: force-disp}, we see here that the detachment process becomes smoother with the rise of the temperature.}
%	\label{fig: xi-disp}
%\end{figure}

Consider an Euler-Bernoulli beam of the sort described in the previous example, but instead of a lateral distributed force, the non-fixed end of the beam is supported at a height $ \bar{u} $. Thus,
\begin{equation*}
    V = \qty{u \Big| u\in H^{2}(\Omega), u(0)=0, u_x(0)=0, u(L)=\bar{u}}.
\end{equation*}
The beam is adhered to a rigid substrate as illustrated in the inset of figure \ref{fig: force-disp}. The adhesion is modeled by a set of $ N $ bonds connected at points $ x_A\in\Omega $ ($ A=1,...,N) $ along the beam. When the $A$-th bond is connected it acts as a linear spring of stiffness $ k_A $, and when it is broken it exerts no force. Hence we use the following potential function to describe the adhesion
\begin{equation}
	\psi_A[u] = \begin{dcases*}
		\frac{1}{2}k_A(u(x_A))^2 &connected\\
		\frac{1}{2}k_AU_A^2 &broken
	\end{dcases*}.
\end{equation}
Here $ U_A $ is constant of units length that describes the broken state potential in terms of elongation of the spring. Note that due to the stochastic nature of our system, each bond may break and reconnect randomly.
%\red{need not break when its elongation reaches $U_A$, rather it may remain in a connected state and break and reconnect again randomly}. 
Accordingly, the state of each bond, either connected or broken, is identified by a two-state \emph{spin variable}. A similar, yet simpler, adhesion-decohesion model was introduced and discussed by Florio et al. \cite{puglisi}. There, it was suggested to introduce a single $N$-state spin variable, $ \xi $, for the entire array of bonds. This is based on the assumption that due to the one-sided decohesion process we have $ \xi $ connected bond at the fixed-end-side and $N-\xi$ broken bonds at the supported-end-side. 
The potential energy of the system is therefore
\begin{equation}
	E[u;\xi] = \int_{\Omega} \frac{1}{2}K_B u_{xx}^2\dd{\Omega} + \sum_{A=0}^{\xi}\frac{1}{2}k\qty(u(x_A))^2 + \frac{1}{2}(N-\xi)kU^2
\end{equation}
%\begin{equation}
%	\tilde{E}[\tilde{u};\xi] = \int_{\tilde{\Omega}} \frac{1}{2}K_B\tilde{u}_{\tilde{x}\tilde{x}}^{2}\dd{\tilde{\Omega}} + \sum_{A=0}^{\xi}\frac{1}{2}k\tilde{u}^2(\tilde{x}_A) + \frac{1}{2}(N-\xi)kU^2
%\end{equation}
%Define the bending energy scale as $ \tilde{E}_0 = K_BU^2/L^3 $, and scale the lengths by $ x=\tilde{x}/L $ and $ u=\tilde{u}/U $; also, define the non-dimensional parameter $ \eta = kU^2/\tilde{E}_0 $ that describes the ratio between adhesion and bending energies. The non-dimensional energy functional hence becomes
%\begin{equation}
%	E[u;\xi] = \int_{\Omega} \frac{1}{2} u_{xx}^{2} \dd{\Omega} + \sum_{A=0}^{\xi}\frac{1}{2}\eta u^2(x_A) + \frac{1}{2}\eta(N-\xi)
%\end{equation}
%where $ \Omega = [0,1] $.
The next step is to approximate $ u $ using Hermite cubic shape functions as was done in the previous example. In principle, one may use any mesh as long as it has nodes at all the $ \qty{x_A} $ points where the beam is attached to the substrate through a breakable bond. For simplicity, we consider here a mesh with $ N $ unknowns located where the springs are connected.
The approximate energy function is then
%\begin{multline}
%	E^h(\vb{d};\xi) = \frac{1}{2}\vb{d}^T\qty(\vb{K}+\vb{K}_{\RR{conc.}}(\xi))\vb{d} \\+ \vb{v}^T\vb{d} + \frac{1}{2}\qty(S + (N-\xi)\eta)
%\end{multline}
\begin{equation}
E^h(\vb{d};\xi) = \frac{1}{2}\vb{d}^T\qty(\vb{K}+\vb{K}_{\RR{conc.}}(\xi))\vb{d} + \vb{v}^T\vb{d} + \frac{1}{2}\qty(S + (N-\xi)\eta)
\end{equation}
where $ \vb{K} $, $ \vb{v} $ and $ S $ are defined as usual, and $ \vb{K}_{\RR{conc.}} $ is a matrix full of zeros except for $ kU^2 $ at $ \xi $ entries along the diagonal corresponding to connected bonds.
The statistical distribution of the system is $ p[u;\xi] = \exp(-\beta E[u;\xi])/Z $, and the discretized version is the well-studied Gaussian distribution.
Once we have $ p^h(\vb{d},\xi) $ we essentially \emph{know everything} about the system, and we can calculate some interesting statistical properties and study how they are influenced by temperature. For example, figure \ref{fig: force-disp}(a) shows the effect of temperature on the force-displacement relation of the mean force, $ \mean{f} = -(\pdv*{Z^h}{\bar{u}})/\beta $, applied by the support with respect to the prescribed displacement $ \bar{u} $.
We may also calculate $ \mean{\xi} $, the mean number of connected bonds, as a function of $ \bar{u} $, as shown in figure \ref{fig: force-disp}(b).
These results are given in \emph{non-dimensional} form after the energy was rescaled by $E_0=K_BU^2/L^3$ and lengths were rescaled by $ U $; the calculations were carried out with the values: $kU^2/E_0 = 5$, $\beta = \qty{15,10,6,4,3,2,1}E_0$ and $N=6$.

\paragraph{Discussion and conclusions.---}
%\emph{Discussion and conclusions$.-$}
We presented a new method for calculating functional integrals based on finite-elements formulation. The proposed method is far more robust, versatile and powerful than any prevailing method, as it allows the calculation of functional integrals over any domain subjected to any boundary conditions or constraints, while not limited to 1-D domains like the “slicing method”. Due to the nature of the discretization, a finer mesh may be used in regions where high accuracy is needed. Moreover, by employing the FE formulation, the functional identity of the state-function is naturally maintained throughout the calculation, enabling insightful perspectives even in 1-D. Just as importantly, existing finite-element routines, elements libraries and shape functions, which have been developed throughout the years for solving PDEs, can be directly employed for calculating functional integrals as well. Three illustrative examples have been discussed, demonstrating the formulation for single and multiple nodal DOFs and showing that the formulation can be conveniently applied to complex systems, even with non-linear behavior. All in all, it is evident that the powerful FE formulation, which revolutionized the numerical analysis of PDEs, combined with modern computing power opens a door for new research opportunities by enabling the study of new problems which could have not been addressed before. Finally, as a secondary effect, the method is likely to accelerate the development and incorporation of new functional-integration schemes as independent modules in existing open-source and commercial FE software.
\\

This work was supported by the Israel Science Foundation (grant No. 1598/21)

%%%%%%%%%%%%%%%
% Bibliography
%%%%%%%%%%%%%%%
%\newpage
\bibliographystyle{apsrev4-2}
\bibliography{FuncInt}

\begin{thebibliography}{23}%
\makeatletter
\providecommand \@ifxundefined [1]{%
 \@ifx{#1\undefined}
}%
\providecommand \@ifnum [1]{%
 \ifnum #1\expandafter \@firstoftwo
 \else \expandafter \@secondoftwo
 \fi
}%
\providecommand \@ifx [1]{%
 \ifx #1\expandafter \@firstoftwo
 \else \expandafter \@secondoftwo
 \fi
}%
\providecommand \natexlab [1]{#1}%
\providecommand \enquote  [1]{``#1''}%
\providecommand \bibnamefont  [1]{#1}%
\providecommand \bibfnamefont [1]{#1}%
\providecommand \citenamefont [1]{#1}%
\providecommand \href@noop [0]{\@secondoftwo}%
\providecommand \href [0]{\begingroup \@sanitize@url \@href}%
\providecommand \@href[1]{\@@startlink{#1}\@@href}%
\providecommand \@@href[1]{\endgroup#1\@@endlink}%
\providecommand \@sanitize@url [0]{\catcode `\\12\catcode `\$12\catcode
  `\&12\catcode `\#12\catcode `\^12\catcode `\_12\catcode `\%12\relax}%
\providecommand \@@startlink[1]{}%
\providecommand \@@endlink[0]{}%
\providecommand \url  [0]{\begingroup\@sanitize@url \@url }%
\providecommand \@url [1]{\endgroup\@href {#1}{\urlprefix }}%
\providecommand \urlprefix  [0]{URL }%
\providecommand \Eprint [0]{\href }%
\providecommand \doibase [0]{https://doi.org/}%
\providecommand \selectlanguage [0]{\@gobble}%
\providecommand \bibinfo  [0]{\@secondoftwo}%
\providecommand \bibfield  [0]{\@secondoftwo}%
\providecommand \translation [1]{[#1]}%
\providecommand \BibitemOpen [0]{}%
\providecommand \bibitemStop [0]{}%
\providecommand \bibitemNoStop [0]{.\EOS\space}%
\providecommand \EOS [0]{\spacefactor3000\relax}%
\providecommand \BibitemShut  [1]{\csname bibitem#1\endcsname}%
\let\auto@bib@innerbib\@empty
%</preamble>
\bibitem [{\citenamefont {Feynman}(1942)}]{Feynman1942}%
  \BibitemOpen
  \bibfield  {author} {\bibinfo {author} {\bibfnamefont {R.~P.}\ \bibnamefont
  {Feynman}},\ }\emph {\bibinfo {title} {The Principle of Least Action in
  Quantum Mechanics}},\ \href {https://doi.org/10.1142/9789812567635_0001}
  {Ph.D. thesis},\ \bibinfo  {school} {Princeton U.} (\bibinfo {year}
  {1942})\BibitemShut {NoStop}%
\bibitem [{\citenamefont {Feynman}(1948)}]{feynman1948}%
  \BibitemOpen
  \bibfield  {author} {\bibinfo {author} {\bibfnamefont {R.~P.}\ \bibnamefont
  {Feynman}},\ }\href {https://doi.org/10.1103/RevModPhys.20.367} {\bibfield
  {journal} {\bibinfo  {journal} {Reviews of Modern Physics}\ }\textbf
  {\bibinfo {volume} {20}},\ \bibinfo {pages} {367} (\bibinfo {year}
  {1948})}\BibitemShut {NoStop}%
\bibitem [{\citenamefont {Feynman}\ and\ \citenamefont
  {Hibbs}(1965)}]{FeynmannHibbs1965}%
  \BibitemOpen
  \bibfield  {author} {\bibinfo {author} {\bibfnamefont {R.~P.}\ \bibnamefont
  {Feynman}}\ and\ \bibinfo {author} {\bibfnamefont {A.~R.}\ \bibnamefont
  {Hibbs}},\ }\href@noop {} {\emph {\bibinfo {title} {Quantum {{Mechanics}} and
  {{Path Integrals}}}}}\ (\bibinfo  {publisher} {{McGraw-Hill}},\ \bibinfo
  {year} {1965})\BibitemShut {NoStop}%
\bibitem [{\citenamefont {Phillips}\ \emph {et~al.}(1998)\citenamefont
  {Phillips}, \citenamefont {Kondev}, \citenamefont {Theriot}, \citenamefont
  {Garcia},\ and\ \citenamefont {Orme}}]{pbcChap10}%
  \BibitemOpen
  \bibfield  {author} {\bibinfo {author} {\bibfnamefont {R.}~\bibnamefont
  {Phillips}}, \bibinfo {author} {\bibfnamefont {J.}~\bibnamefont {Kondev}},
  \bibinfo {author} {\bibfnamefont {J.}~\bibnamefont {Theriot}}, \bibinfo
  {author} {\bibfnamefont {H.~G.}\ \bibnamefont {Garcia}},\ and\ \bibinfo
  {author} {\bibfnamefont {N.}~\bibnamefont {Orme}},\ }in\ \href@noop {} {\emph
  {\bibinfo {booktitle} {Physical {{Biology}} of the {{Cell}}}}}\ (\bibinfo
  {publisher} {{Garland Science}},\ \bibinfo {year} {1998})\ \bibinfo {edition}
  {2nd}\ ed.\BibitemShut {Stop}%
\bibitem [{\citenamefont {Wang}\ \emph {et~al.}(2006)\citenamefont {Wang},
  \citenamefont {Zhang}, \citenamefont {Lu},\ and\ \citenamefont
  {Wang}}]{chemistryPRL}%
  \BibitemOpen
  \bibfield  {author} {\bibinfo {author} {\bibfnamefont {J.}~\bibnamefont
  {Wang}}, \bibinfo {author} {\bibfnamefont {K.}~\bibnamefont {Zhang}},
  \bibinfo {author} {\bibfnamefont {H.}~\bibnamefont {Lu}},\ and\ \bibinfo
  {author} {\bibfnamefont {E.}~\bibnamefont {Wang}},\ }\href
  {https://doi.org/10.1103/PhysRevLett.96.168101} {\bibfield  {journal}
  {\bibinfo  {journal} {Physical Review Letters}\ }\textbf {\bibinfo {volume}
  {96}},\ \bibinfo {pages} {168101} (\bibinfo {year} {2006})}\BibitemShut
  {NoStop}%
\bibitem [{\citenamefont {Wang}\ \emph {et~al.}(2010)\citenamefont {Wang},
  \citenamefont {Zhang},\ and\ \citenamefont {Wang}}]{chemistryJCP}%
  \BibitemOpen
  \bibfield  {author} {\bibinfo {author} {\bibfnamefont {J.}~\bibnamefont
  {Wang}}, \bibinfo {author} {\bibfnamefont {K.}~\bibnamefont {Zhang}},\ and\
  \bibinfo {author} {\bibfnamefont {E.}~\bibnamefont {Wang}},\ }\href
  {https://doi.org/10.1063/1.3478547} {\bibfield  {journal} {\bibinfo
  {journal} {The Journal of Chemical Physics}\ }\textbf {\bibinfo {volume}
  {133}},\ \bibinfo {pages} {125103} (\bibinfo {year} {2010})}\BibitemShut
  {NoStop}%
\bibitem [{\citenamefont {Zan}\ \emph {et~al.}(2022)\citenamefont {Zan},
  \citenamefont {Jia},\ and\ \citenamefont {Xu}}]{Reliability1}%
  \BibitemOpen
  \bibfield  {author} {\bibinfo {author} {\bibfnamefont {W.}~\bibnamefont
  {Zan}}, \bibinfo {author} {\bibfnamefont {W.}~\bibnamefont {Jia}},\ and\
  \bibinfo {author} {\bibfnamefont {Y.}~\bibnamefont {Xu}},\ }\href
  {https://doi.org/10.1016/j.probengmech.2022.103252} {\bibfield  {journal}
  {\bibinfo  {journal} {Probabilistic Engineering Mechanics}\ }\textbf
  {\bibinfo {volume} {68}},\ \bibinfo {pages} {103252} (\bibinfo {year}
  {2022})}\BibitemShut {NoStop}%
\bibitem [{\citenamefont {Naess}\ \emph {et~al.}(2011)\citenamefont {Naess},
  \citenamefont {Iourtchenko},\ and\ \citenamefont {Batsevych}}]{Reliability2}%
  \BibitemOpen
  \bibfield  {author} {\bibinfo {author} {\bibfnamefont {A.}~\bibnamefont
  {Naess}}, \bibinfo {author} {\bibfnamefont {D.}~\bibnamefont {Iourtchenko}},\
  and\ \bibinfo {author} {\bibfnamefont {O.}~\bibnamefont {Batsevych}},\ }\href
  {https://doi.org/10.1016/j.probengmech.2010.05.005} {\bibfield  {journal}
  {\bibinfo  {journal} {Probabilistic Engineering Mechanics}\ }\bibinfo
  {series} {Special {{Issue}}: {{Stochastic Methods}} in {{Mechanics}}
  \textemdash{} {{Status}} and {{Challenges}}},\ \textbf {\bibinfo {volume}
  {26}},\ \bibinfo {pages} {5} (\bibinfo {year} {2011})}\BibitemShut {NoStop}%
\bibitem [{\citenamefont {Linetsky}(1998)}]{finance}%
  \BibitemOpen
  \bibfield  {author} {\bibinfo {author} {\bibfnamefont {V.}~\bibnamefont
  {Linetsky}},\ }\href@noop {} {\bibfield  {journal} {\bibinfo  {journal}
  {Kluwar Academic Publisher}\ ,\ \bibinfo {pages} {35}} (\bibinfo {year}
  {1998})}\BibitemShut {NoStop}%
\bibitem [{\citenamefont {Hughes}(2000)}]{hughes}%
  \BibitemOpen
  \bibfield  {author} {\bibinfo {author} {\bibfnamefont {T.~J.~R.}\
  \bibnamefont {Hughes}},\ }\href@noop {} {\emph {\bibinfo {title} {The
  {{Finite Element Method}}: {{Linear Static}} and {{Dynamic Finite Element
  Analysis}}}}}\ (\bibinfo  {publisher} {{Dover Publications}},\ \bibinfo
  {year} {2000})\BibitemShut {NoStop}%
\bibitem [{\citenamefont {Bathe}(1996)}]{bathe}%
  \BibitemOpen
  \bibfield  {author} {\bibinfo {author} {\bibfnamefont {K.~J.}\ \bibnamefont
  {Bathe}},\ }\href@noop {} {\emph {\bibinfo {title} {Finite {{Element
  Procedures}}}}}\ (\bibinfo  {publisher} {{Prentice Hall}},\ \bibinfo {year}
  {1996})\BibitemShut {NoStop}%
\bibitem [{\citenamefont {Bastos}\ and\ \citenamefont
  {Sadowski}(2017)}]{magneticFE}%
  \BibitemOpen
  \bibfield  {author} {\bibinfo {author} {\bibfnamefont {J.~P.~A.}\
  \bibnamefont {Bastos}}\ and\ \bibinfo {author} {\bibfnamefont
  {N.}~\bibnamefont {Sadowski}},\ }\href {https://doi.org/10.1201/b15558}
  {\emph {\bibinfo {title} {Magnetic {{Materials}} and {{3D Finite Element
  Modeling}}}}}\ (\bibinfo  {publisher} {{CRC Press}},\ \bibinfo {year}
  {2017})\BibitemShut {NoStop}%
\bibitem [{\citenamefont {Volakis}\ \emph {et~al.}(1994)\citenamefont
  {Volakis}, \citenamefont {Chatterjee},\ and\ \citenamefont
  {Kempel}}]{scattering}%
  \BibitemOpen
  \bibfield  {author} {\bibinfo {author} {\bibfnamefont {J.~L.}\ \bibnamefont
  {Volakis}}, \bibinfo {author} {\bibfnamefont {A.}~\bibnamefont
  {Chatterjee}},\ and\ \bibinfo {author} {\bibfnamefont {L.~C.}\ \bibnamefont
  {Kempel}},\ }\href {https://doi.org/10.1364/JOSAA.11.001422} {\bibfield
  {journal} {\bibinfo  {journal} {JOSA A}\ }\textbf {\bibinfo {volume} {11}},\
  \bibinfo {pages} {1422} (\bibinfo {year} {1994})}\BibitemShut {NoStop}%
\bibitem [{\citenamefont {Andonegui}\ and\ \citenamefont
  {{Garcia-Adeva}}(2013)}]{photonicCrystals}%
  \BibitemOpen
  \bibfield  {author} {\bibinfo {author} {\bibfnamefont {I.}~\bibnamefont
  {Andonegui}}\ and\ \bibinfo {author} {\bibfnamefont {A.~J.}\ \bibnamefont
  {{Garcia-Adeva}}},\ }\href {https://doi.org/10.1364/OE.21.004072} {\bibfield
  {journal} {\bibinfo  {journal} {Optics Express}\ }\textbf {\bibinfo {volume}
  {21}},\ \bibinfo {pages} {4072} (\bibinfo {year} {2013})}\BibitemShut
  {NoStop}%
\bibitem [{\citenamefont {Zienkiewicz}\ \emph {et~al.}(2000)\citenamefont
  {Zienkiewicz}, \citenamefont {Taylor},\ and\ \citenamefont
  {Taylor}}]{zienkiewiczFE}%
  \BibitemOpen
  \bibfield  {author} {\bibinfo {author} {\bibfnamefont {O.~C.}\ \bibnamefont
  {Zienkiewicz}}, \bibinfo {author} {\bibfnamefont {R.~L.}\ \bibnamefont
  {Taylor}},\ and\ \bibinfo {author} {\bibfnamefont {R.~L.}\ \bibnamefont
  {Taylor}},\ }\href@noop {} {\emph {\bibinfo {title} {The {{Finite Element
  Method}}: {{Solid}} Mechanics}}}\ (\bibinfo  {publisher}
  {{Butterworth-Heinemann}},\ \bibinfo {year} {2000})\BibitemShut {NoStop}%
\bibitem [{\citenamefont {White}(1985)}]{whiteFE}%
  \BibitemOpen
  \bibfield  {author} {\bibinfo {author} {\bibfnamefont {R.~E.}\ \bibnamefont
  {White}},\ }\href@noop {} {\emph {\bibinfo {title} {An {{Introduction}} to
  the {{Finite Element Method}} with {{Applications}} to {{Nonlinear
  Problems}}}}}\ (\bibinfo  {publisher} {{Wiley}},\ \bibinfo {year}
  {1985})\BibitemShut {NoStop}%
\bibitem [{\citenamefont {Rao}(2005)}]{raoFE}%
  \BibitemOpen
  \bibfield  {author} {\bibinfo {author} {\bibfnamefont {S.~S.}\ \bibnamefont
  {Rao}},\ }\href@noop {} {\emph {\bibinfo {title} {The {{Finite Element
  Method}} in {{Engineering}}}}}\ (\bibinfo  {publisher}
  {{Butterworth-Heinemann}},\ \bibinfo {year} {2005})\BibitemShut {NoStop}%
\bibitem [{\citenamefont {Belytschko}\ \emph {et~al.}(2009)\citenamefont
  {Belytschko}, \citenamefont {Gracie},\ and\ \citenamefont
  {Ventura}}]{fractureFE}%
  \BibitemOpen
  \bibfield  {author} {\bibinfo {author} {\bibfnamefont {T.}~\bibnamefont
  {Belytschko}}, \bibinfo {author} {\bibfnamefont {R.}~\bibnamefont {Gracie}},\
  and\ \bibinfo {author} {\bibfnamefont {G.}~\bibnamefont {Ventura}},\ }\href
  {https://doi.org/10.1088/0965-0393/17/4/043001} {\bibfield  {journal}
  {\bibinfo  {journal} {Modelling and Simulation in Materials Science and
  Engineering}\ }\textbf {\bibinfo {volume} {17}},\ \bibinfo {pages} {043001}
  (\bibinfo {year} {2009})}\BibitemShut {NoStop}%
\bibitem [{\citenamefont {{The DefElement contributors}}(2023)}]{defelement}%
  \BibitemOpen
  \bibfield  {author} {\bibinfo {author} {\bibnamefont {{The DefElement
  contributors}}},\ }\href@noop {} {\bibinfo {title}
  {{{DefElement}}:~an~encyclopedia~of~finite~element~definitions}},\ \bibinfo
  {howpublished} {https://defelement.com} (\bibinfo {year} {2023})\BibitemShut
  {NoStop}%
\bibitem [{\citenamefont {Geuzaine}\ and\ \citenamefont
  {Remacle}(2023)}]{gmsh}%
  \BibitemOpen
  \bibfield  {author} {\bibinfo {author} {\bibfnamefont {C.}~\bibnamefont
  {Geuzaine}}\ and\ \bibinfo {author} {\bibfnamefont {J.-F.}\ \bibnamefont
  {Remacle}},\ }\href@noop {} {\bibinfo {title} {Gmsh}} (\bibinfo {year}
  {2023})\BibitemShut {NoStop}%
\bibitem [{\citenamefont {Dunn}(2012)}]{Dunn}%
  \BibitemOpen
  \bibfield  {author} {\bibinfo {author} {\bibfnamefont {W.~L.}\ \bibnamefont
  {Dunn}},\ }\href@noop {} {\emph {\bibinfo {title} {Exploring {{Monte Carlo}}
  Methods}}}\ (\bibinfo  {publisher} {{Elsevier}},\ \bibinfo {address}
  {{Amsterdam}},\ \bibinfo {year} {2012})\BibitemShut {NoStop}%
\bibitem [{\citenamefont {Reif}(2009)}]{reif}%
  \BibitemOpen
  \bibfield  {author} {\bibinfo {author} {\bibfnamefont {F.}~\bibnamefont
  {Reif}},\ }\href@noop {} {\emph {\bibinfo {title} {Fundamentals of
  {{Statistical}} and {{Thermal Physics}}}}}\ (\bibinfo  {publisher} {{Waveland
  Press}},\ \bibinfo {year} {2009})\BibitemShut {NoStop}%
\bibitem [{\citenamefont {Florio}\ \emph {et~al.}(2020)\citenamefont {Florio},
  \citenamefont {Puglisi},\ and\ \citenamefont {Giordano}}]{puglisi}%
  \BibitemOpen
  \bibfield  {author} {\bibinfo {author} {\bibfnamefont {G.}~\bibnamefont
  {Florio}}, \bibinfo {author} {\bibfnamefont {G.}~\bibnamefont {Puglisi}},\
  and\ \bibinfo {author} {\bibfnamefont {S.}~\bibnamefont {Giordano}},\ }\href
  {https://doi.org/10.1103/PhysRevResearch.2.033227} {\bibfield  {journal}
  {\bibinfo  {journal} {Physical Review Research}\ }\textbf {\bibinfo {volume}
  {2}},\ \bibinfo {pages} {033227} (\bibinfo {year} {2020})}\BibitemShut
  {NoStop}%
\end{thebibliography}%

%%%%%%%%%%%%%%%
% Appendix
%%%%%%%%%%%%%%%
%\newpage
%\appendix
\end{document}